\documentclass[final,5p,times,twocolumn]{elsarticle}

 \usepackage{graphics}
 \usepackage{graphicx}
 
\usepackage{amssymb}
\usepackage{amsthm}

\usepackage{bbold}

\usepackage{algorithm}
\usepackage{algorithmicx}
\usepackage{algpseudocode}

\makeatletter
\def\ps@pprintTitle{%
  \let\@oddhead\@empty
  \let\@evenhead\@empty
  \def\@oddfoot{\reset@font\hfil\thepage\hfil}
  \let\@evenfoot\@oddfoot
}
\makeatother

\journal{}

\begin{document}

\makeatletter
\newlength \figwidth
\if@twocolumn
  \setlength \figwidth {1\columnwidth}
\else
  \setlength \figwidth {0.8\textwidth}
\fi
\makeatother

\begin{frontmatter}




\title{Effects of Individual Success on Globally Distributed Team Performance}


\author{Onur Y{\i}lmaz }
\address{Computer Engineering Department, Middle East Technical University, Ankara, Turkey}
\ead{<onur@onuryilmaz.me>}


\begin{abstract}

Necessity of different competencies with high level of knowledge makes it inevitable that software development is a team work. With the today’s technology, teams can communicate both synchronously and asynchronously using different online collaboration tools throughout the world. Researches indicate that there are many factors that affect the team success and in this paper, effect of individual success on globally distributed team performance will be analyzed. Student team projects undertaken by other researchers will be used to analyze collected data and conclusions will be drawn for further analysis.

\end{abstract}


\end{frontmatter}


\section{Introduction}
\label{}

Considering the necessity and variety of competency, it is inevitable that software development is a cooperative work. In other words, software engineering is based on group activity where each player is assigned with the role related to different responsibilities. Although, teamwork is thought as a must, most of the project failures are related to the deficiency of team configurations. Therefore, having the optimum team composition is crucial for software development projects \cite{1_khaled17building}.

With the tools available online, in today’s world, collaboration can be conducted throughout the world without any face-to-face communication obligation. When these applications are considered, both synchronous and asynchronous applications are available for online collaboration. For synchronous applications, video chat and instant messaging can be the most progressive examples as of today, whereas asynchronous tools can be illustrated with email, social networking and forums. With the help of these tools, people from different universities, countries and even from different continents can work on the same topic with a high degree of collaboration  \cite{2_DBLP:journals/cacm/Meyer13}. 

Grouping people from different backgrounds and regions have very common drawbacks. Considering different time-zones, communication within teams and managing them is a rigorous work. In addition, people from different programming backgrounds result with large range of experience and knowledge, which are indispensable for software development. Therefore, software development teams must be considered as “an optimum combination of competencies” rather than just a group of developers.

As mentioned, software development needs different responsibilities and competencies, which can be gathered together from different regions of the world by means of today’s technology. In addition, there are many factors that can affect the performance of a global software development team, such as cultural, individual and collaborative group work attitude \cite{3_swigger2009structural}. In this study, a subclass of individual factors, namely individual success, will be analyzed in the sense of affecting the team performance. Individual success of the team players will be based on their experience and GPA; on the other hand team performance will be considered in two dimensions as overall grade and team communication. Although reasoning of analyzing overall team grade is self-explanatory, team communication analysis will also be undertaken because research on software teams indicate the fact that team communication is affected by frequency of communication within teams  \cite{4_Dutoit:1998:CMS:286181.286185}. 

In this paper, projects which are conducted by previous researchers in this area are analyzed. Software development teams studied in this research were group of students from different universities and countries who are assigned an assignment project and different communication tools are provided for them. Goal of this research is extracting underlying relationships between the individual success of people and the successful collaborations in globally distributed teams. Analysis resulted with some important findings. Firstly, it could be stated that higher average academic success of team members yield more successful teams. Secondly, as range of team members’ success increases, a decrease in team performance is observed. Thirdly and finally, as GPA of team members’ increase, their contribution to team communication is increasing, which is an important driver of success in this environment.

\section{Relevant Research}
\label{}

\subsection{Individual Success}
Individual characteristics like academic success, age, sex, education level and experience contribute on team performance in collaborative teams \cite{3_swigger2009structural}. In addition, some researches indicate that with the increasing range of individual characteristics within group leads to decrease in team performance \cite{5_bochner1994power}. When the scope of individual characteristics is limited to individual success, it can be concluded that one person’s educational background and work experience can considerably affect the perception on his/her work \cite{6_4095018}.

\subsection{Globally Distributed Teams}
Although some of the drawbacks of globally distributed teams are mentioned, both industry and institutions continue to work with small groups which are combination of people from different regions \cite{5_bochner1994power}. Researches also mention that student teams and professionals cannot be always analyzed with the same reasoning, therefore the conclusions inferred from researches based on student teams should not be followed in industry without delicate analysis \cite{3_swigger2009structural}.

\subsection{Team Performance}
Since the projects in this research based on capabilities provided by online collaboration, effect of communication on team performance should be considered. Research about this issue reveals that frequency of communication affects the performance of the team \cite{3_swigger2009structural}. Because communication frequency shows the number of times team members interacted each other, the mentioned result is not surprising but detailed analysis should also be undertaken to check the relationships and underlying effects.

\section{Methodology}
\label{}

\subsection{Overall Design of the Study}
In this research, two independent student team projects which are conducted in fall and spring semesters of 2009 are used. Students from three universities, namely Atilim University (AU), Universidad Tecnologica de Panama (UTP) and University of North Texas (UNT), are attended to these projects. Scope of their works are based on design and implementation of their assignments. In addition, they are provided with a set of online communication tools and they are trained on how to use them. Usage statistics of these communication tools are collected and students were pre-informed about this. Full coverage of these projects and data collection were in the scope of the article of Serce et al. \cite{7_SerceSABDC11}.

\subsection{Projects and Participants}
As already mentioned, in this study, two projects which are already undertaken by other researchers are used. These projects were studied in the paper of Serce et al. with the name of ``Online Collaboration: Collaborative Behavior Patterns and Factors Affecting Globally Distributed Team Performance'' \cite{7_SerceSABDC11}. Important aspects of these projects related to this study can be summarized as following:

\textbf{Project \#1:} In this project, student teams are assigned with a database management system for car rental agency in the fall semester of 2009 for 6 weeks. Scope of the work is based on design, functionality assessment, implementing and testing. Participants were from Atilim University (AU), Universidad Tecnologica de Panama (UTP) and University of North Texas (UNT).  

\textbf{Project \#2:} In this project, student teams are assigned with a standalone bookstore management application which can be used by bookstore staff for daily operations. Scope of this project was design and implementation of this application and hence the duration was nearly two months. Participating universities are same with the Project \#1 and this project is conducted during the spring semester of 2009. 

Number of students with their universities and their roles in these projects are summarized in Table~\ref{table1} and Table~\ref{table2} as following.

\begin{table}[ht]
\caption{Number of students from each university in projects} 
\centering  
\includegraphics[width=\figwidth]{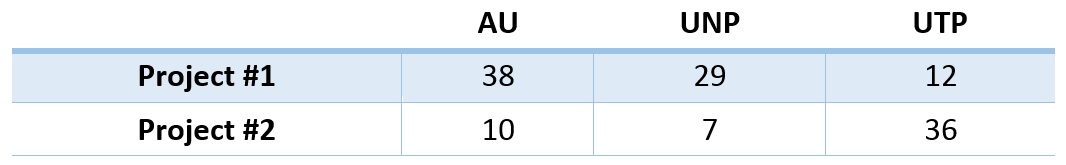}
\label{table1} 
\end{table}

\begin{table}[ht]
\caption{Roles of students from each university in projects} 
\centering  
\includegraphics[width=\figwidth]{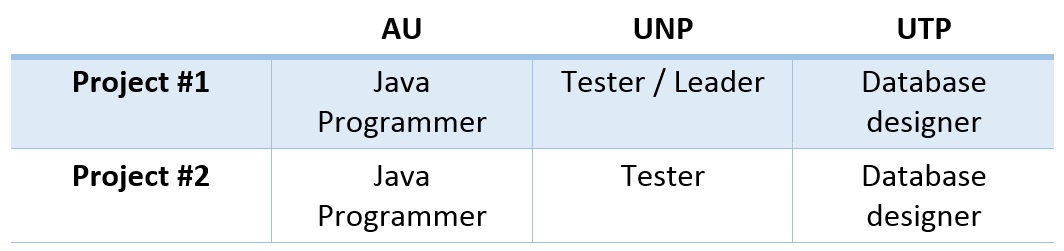}
\label{table2} 
\end{table}

\subsection{Measures and Data Collection}
There are three important aspects of data related to this study and they are GPA for indicator of individual success, team performance grades for measuring team success levels and communication statistics which are directly influencer of team success in this environment \cite{3_swigger2009structural}. Although vast amount of communication related statistics are collected, there are significant level of missing data in GPA values of students which are counted as missing and removed from data completely in analysis stage. This loss data restricted this research conducting comprehensive statistical analysis and thus simple methods are used for correlation and revealing underlying relationships.

\subsection{Data Analysis}

Firstly, in order to present the general situation, GPA values of all participants are analyzed for each project. As tabulated in Table~\ref{table3} below, it can be seen that projects have no significant difference in GPA values in the sense of mean, standard deviation and ranges. In addition, when their histograms are checked from Figure 1 and 2, even there are less number of samples no significant skewness in the graphs is observed.

\begin{table}[ht]
\caption{Analysis of GPA for both projects projects} 
\centering  
\includegraphics[width=\figwidth]{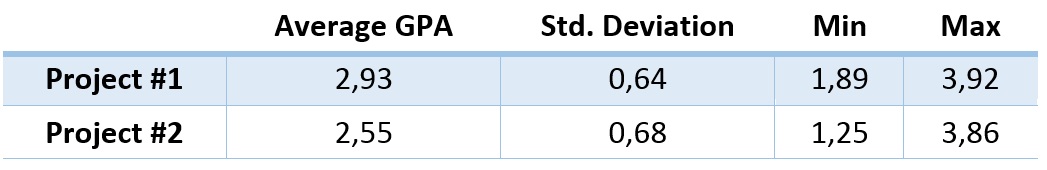}
\label{table3} 
\end{table}

\begin{figure}[h!]
  \includegraphics[width=\figwidth]{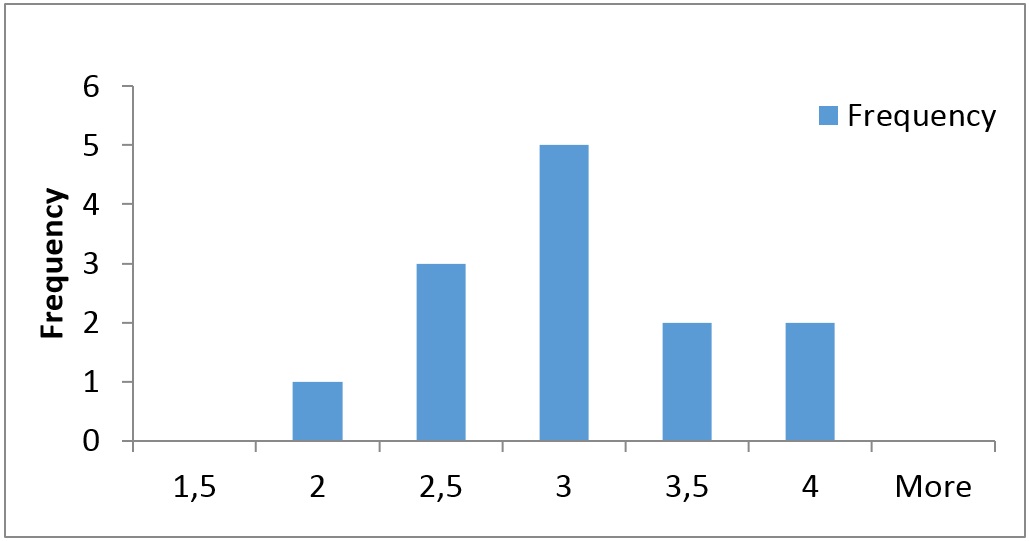}
  \caption{GPA Histogram of Project \#1}
  \label{figure1}
\end{figure}

\begin{figure}[h!]
  \includegraphics[width=\figwidth]{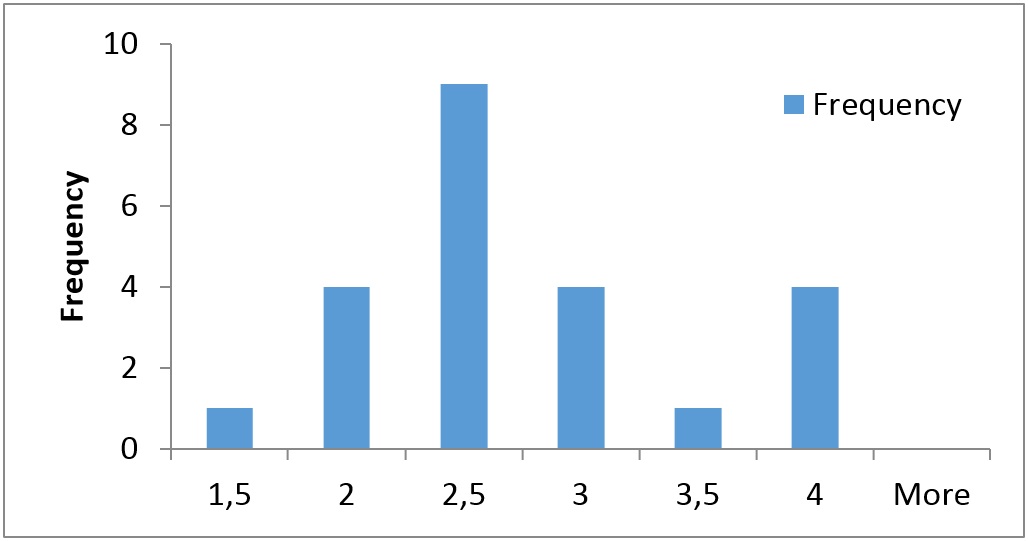}
  \caption{GPA Histogram of Project \#2}
  \label{figure2}
\end{figure}

\begin{figure}[h!]
  \includegraphics[width=\figwidth]{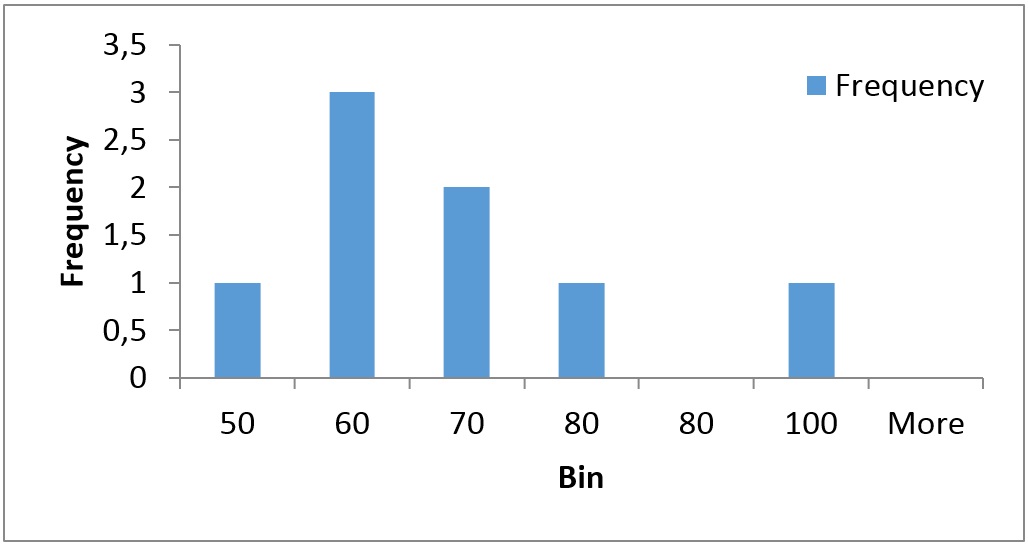}
  \caption{Performance Grade Histogram of Project \#1}
  \label{figure3}
\end{figure}

\begin{figure}[h!]
  \includegraphics[width=\figwidth]{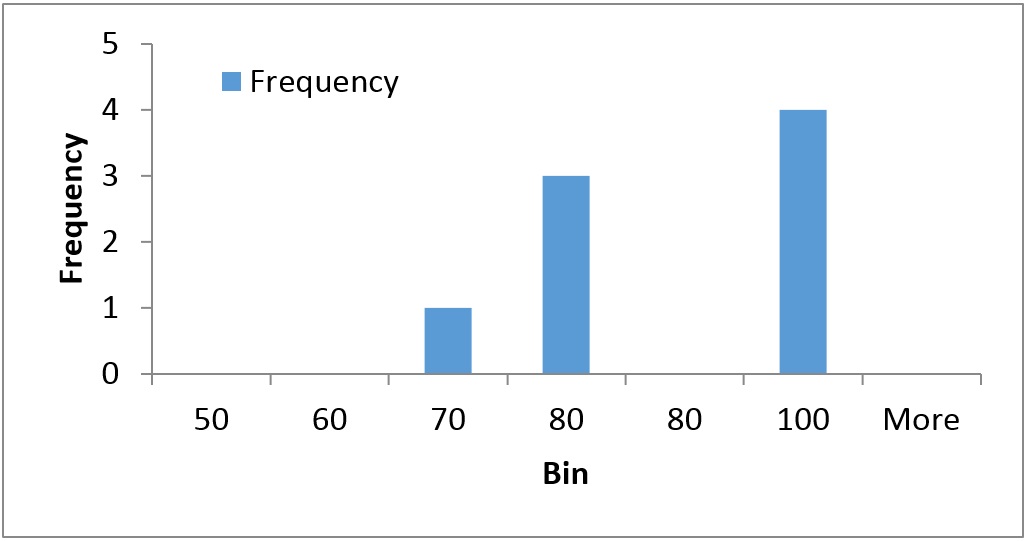}
  \caption{Performance Grade Histogram of Project \#2}
  \label{figure4}
\end{figure}

Secondly, general situation of the team performances is analyzed for both projects. While conducting these projects, all students were assigned with a performance grade and for each team,  team averages are  taken as team  performance  grade  in this study. With the same approach used in the prior step, analysis of team performance values are tabulated in Table~\ref{table4}. It can be seen that, average performance grade and range is slightly shifted in Project \#2.  In addition, when histograms of performance grades are checked from Figure~\ref{figure3} and ~\ref{figure4}, it can be seen that the second project grades are articulated and higher than the first project grades. 

\begin{table}[ht]
\caption{Analysis of performance grades for both projects} 
\centering  
\includegraphics[width=\figwidth]{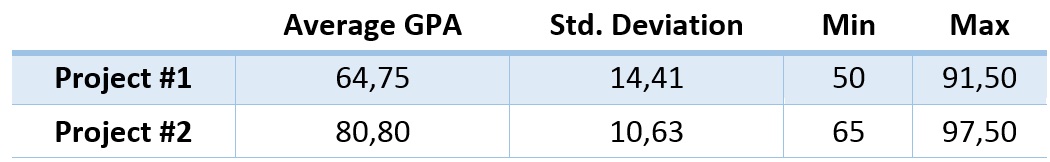}
\label{table4} 
\end{table}

After presenting the individual analysis of measures, thirdly, effect of average GPA of development team on team performance is analyzed. For each team, mean of GPA values are calculated which is an indicator of the general academic success of the team. With the same approach used in the prior analysis, average performance grades are calculated as team performance indicator. Relationship between these measures can be seen from the Figure~\ref{figure5} below. Scatter diagram shows that for both projects, there is a tendency for increase in average team performance as average team GPA increases. Since the second project has higher performance grades in general, effect of average GPA is not standing out as the first project.

\begin{figure}[h!]
  \includegraphics[width=\figwidth]{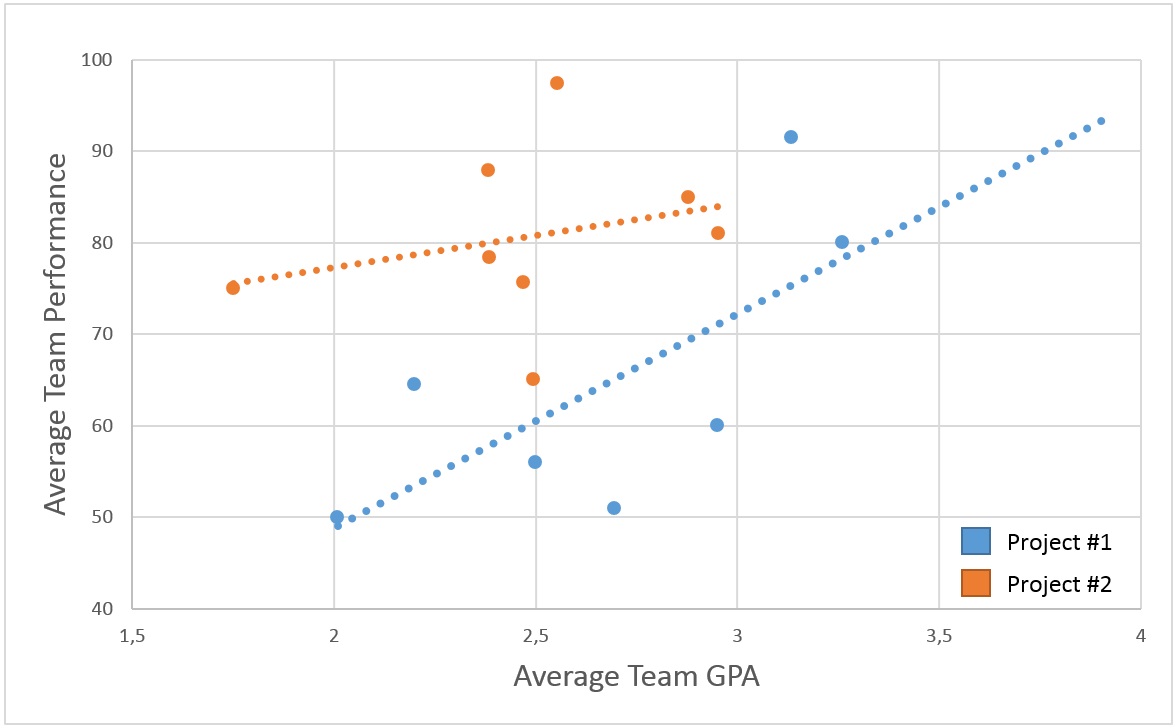}
  \caption{Average Team Performance vs. Average Team GPA}
  \label{figure5}
\end{figure}

\begin{figure}[h!]
  \includegraphics[width=\figwidth]{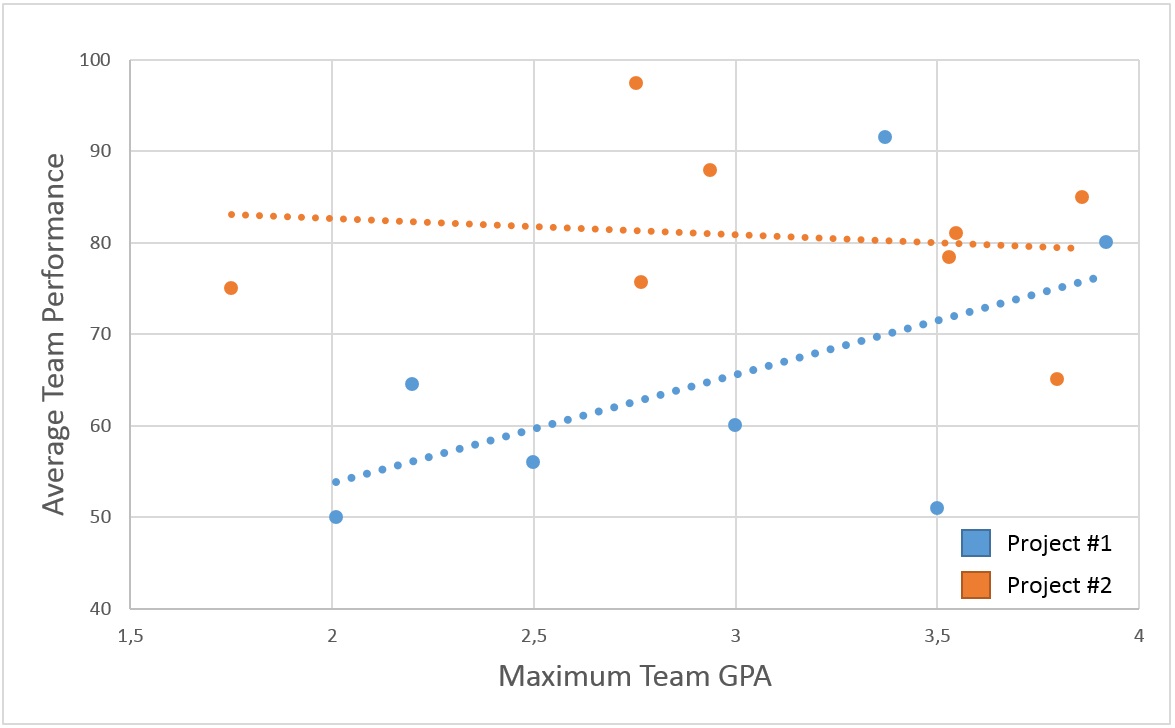}
  \caption{Average Team Performance vs. Maximum Team GPA}
  \label{figure6}
\end{figure}

After the analysis of the average team success, effect of the most successful team player on the average team success is analyzed. With this aim, for each team, student with the highest GPA value is selected and relationship between the average team performances is checked. Although it is expected that having a team member with higher GPA values can push team to have higher grades in general, this situation does not hold for both projects. In other words, where Project \#1 has positive correlation, Project \#2 reveals a negative correlation between these measures as it can be seen from Figure~\ref{figure6}.

As mentioned before, researches indicate difference in team members’ individual characteristics yield decrease in team performance. Considering the focus of this paper, maximum difference in the GPA of team members for each team is calculated. When the relationship between GPA difference within team and average team performance is checked from the Figure~\ref{figure7} below, negative correlation can be easily seen. In other words, for both projects as the GPA range within students increase average team performance decreases. 

\begin{figure}[h!]
  \includegraphics[width=\figwidth]{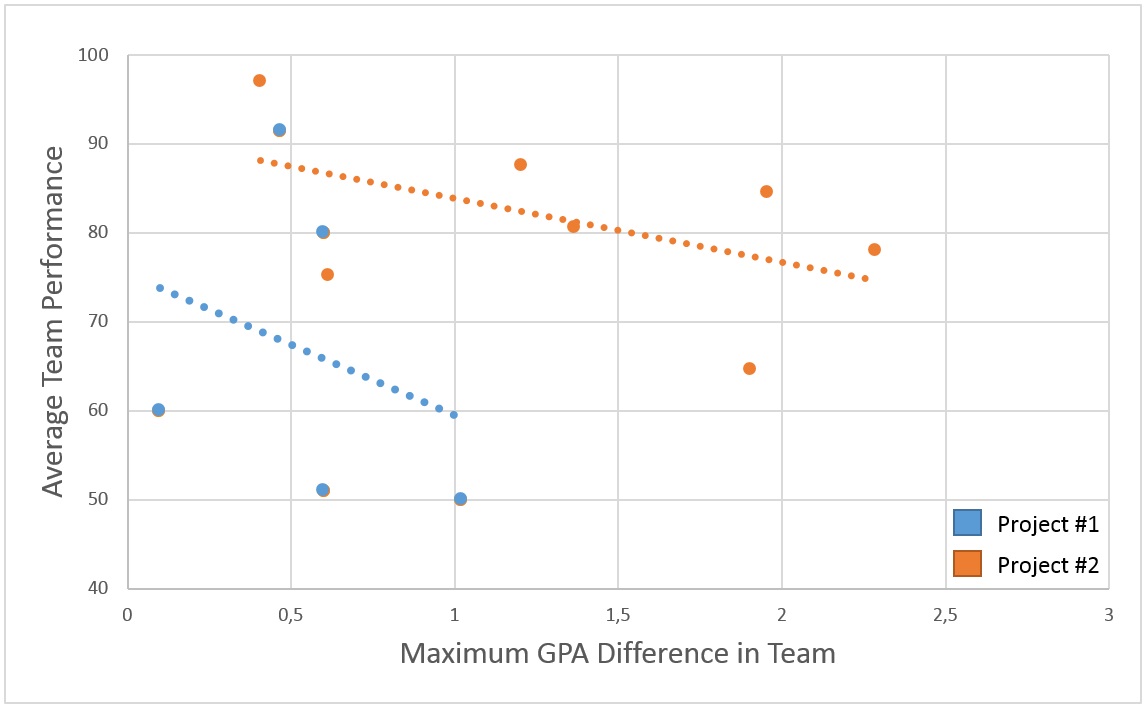}
  \caption{Average Team Performance vs. Maximum GPA Difference in Team}
  \label{figure7}
\end{figure}

\begin{figure}[h!]
  \includegraphics[width=\figwidth]{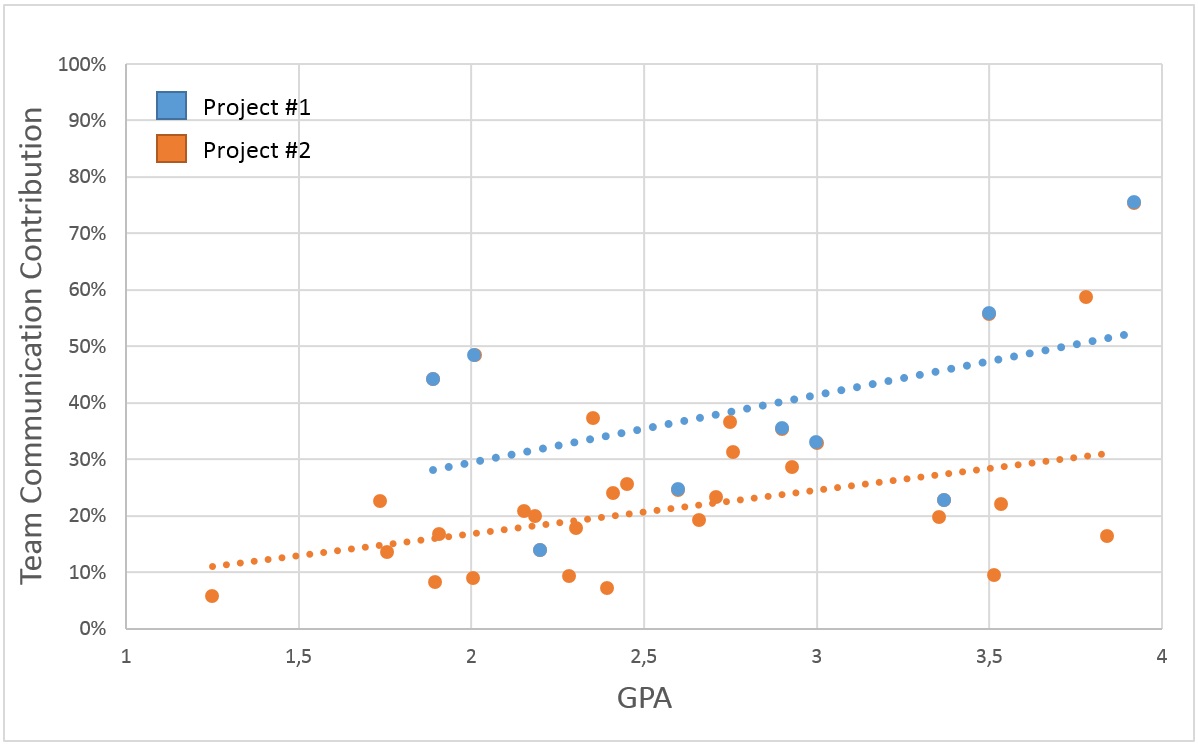}
  \caption{Team Communication Contribution vs. GPA}
  \label{figure8}
\end{figure}

In the relevant research section, it is mentioned that studies reveal the fact that frequency of communication affects the performance of the team \cite{3_swigger2009structural}. Considering this, how GPA of a team member is related to his/her contribution to communication is analyzed. Before going into further analysis, it should be mentioned that there are many different communication tools available for these projects, like forums, email, chat and wiki. However, since analyzing and assessing differences and contributions of each tool is out of scope of this paper, all of them is counted as communication session. For each student, percentage of their contribution to team communication is calculated and its relationship to GPAs are checked. From the scatter diagram in Figure~\ref{figure8}, it can be seen that there is a positive correlation between GPA and contribution to communication for both projects.
 
\section{Discussion}
Data analysis showed that there are different underlying relationship between GPA and team success. Although there is high level of missing data, which restricts undertaking detailed statistical analysis, correlation between measures can easily be detected. Considering the necessity for further studies, some conclusions from these analysis are generalized and drawn:

\begin{itemize} 
\renewcommand{\labelitemi}{$\bullet$}

\item When the average GPA of a team increases, team performance is escalated. This shows that individually successful team members can work in harmony and create a successful teamwork environment, which ultimately yields teams with higher performance.
\item Increasing range of GPA decreases the team performance. This confirms the researches mentioning that the difference in individual characteristics yields decreasing team success. This conclusion is based on the fact that as range increases, there is at least one team member with decreasing GPA. Having unsuccessful students in the teams, it is inevitable that the overall team performance will be falling down.
\item Team members with higher GPA communicate more in teams. This conclusion is based on the fact that, for these projects each type of communication is counted as same without considering their characteristics. Since communication frequency shows the number of interactions, as communication increases it can be concluded that “a work is in progress”. This is confirmed by the two projects, where in each percentage of the team communication increases with the higher GPA of the students.
\item Although team members with higher GPA values are expected to pull their teams to be more successful, for these two projects this statement cannot be verified in this study. This is mostly based on the fact that, only one individual with high GPA cannot affect the whole team’s work attitude and performance grade. This asymmetry in the teams resulted with the fact that no concrete relation between the highest GPA in the team and team performance can be found.
\end{itemize}

 \section{Conclusion and Future Work}
 In this paper, effect of individual success on globally distributed software teams is studied. In order to reveal underlying relationships, student team projects which are conducted by other researchers and their data is used. In these projects, students from Panama, Turkey and the United States participated and their individual characteristics and communication statistics are collected \cite{3_swigger2009structural}. Since data is collected and shared by other researchers, there was a noteworthy level of missing data; however, projects with the most complete dataset are selected and used in this paper. Analysis of the data with the goal of extracting underlying relationships between the individual success of people and the successful collaborations in globally distributed teams is undertaken. In the light of the relevant research and this analysis, some remarkable findings are constructed and presented. 

For future work on this subject, there some important points to mention. Firstly, in this paper a comprehensive statistical analysis is not implemented due to low number of data. However, with a complete and large dataset, statistical analysis and validation should be undertaken to clarify and support findings. Secondly, in this paper student teams are studied but the conclusions provide insight for implementing in the industry. Before implementing these findings in the professional life, delicate analysis of compatibility should be undertaken. Thirdly and finally, dimension of individual success is studied as GPA in this study. However, studies indicate that individual success can be defined on multiple parameters such as work experience level or knowledge on the related subject. Considering this, different parameters of individual success can be studied to reveal their relationships to team performance. With the help of these future studies, the presented conclusions and analysis will be more concrete and drawbacks of them will be overcome.




\section{References}
\label{}

 
\bibliographystyle{elsarticle-num}

\bibliography{references}
 
\end{document}